\documentclass{article}
 
\usepackage{amsmath,amsthm,amssymb}

\usepackage{cite}

\pdfoutput=1

\usepackage{graphics,graphicx}
\usepackage{epsfig}
\usepackage{multicol}
\usepackage{color}
\makeatletter
\@addtoreset{equation}{section}
\makeatother

\usepackage{braket}
\usepackage{bm}
\usepackage{graphicx}

\usepackage{subcaption}
\usepackage{relsize}

\usepackage{soul}

\usepackage{lineno}

\usepackage{authblk}
\usepackage{hyperref}

\begin{document}

\title{Towards a Quantum Erlangen Program}
\author[1,2]{Matthew J. Lake\footnote{matthew.lake@ubbcluj.ro}} 
\affil[1]{Department of Physics, Babe\c s-Bolyai University, Mihail Kog\u alniceanu Street 1, 400084 Cluj-Napoca, Romania}

\maketitle

\begin{abstract}
The classical Erlangen Program sought to classify metric spaces entirely in terms of their symmetries. 
In physical spacetimes, these symmetries define transformations between classical reference frames, yielding a one-to-one correspondence between frame transformations and the underlying geometry. 
More recently, the classical notion of ideal frame has been extended to the quantum regime, by considering observers as embodied physical systems, subject to the laws of quantum mechanics.
Here, we build on this approach, but outline an alternative definition of the term `quantum reference frame', which differs somewhat from the mainstream view. 
We then show how the new definition can be used to construct a simple model of Planck-scale spacetime, which makes contact with existing quantum gravity phenomenology. 
Finally, we show how classical spacetime symmetries can be ``mathematically preserved but operationally broken'', in the new model, suggesting that {\it quantum} spacetime may be classified, at least locally, in terms of transformations between quantised frames of reference.  
\end{abstract}

This work is based on a talk given at the 13th Bolyai-Gauss-Lobachevsky Conference on Non-Euclidean Geometry in Oujda, Morocco, in April 2025.
\\ \\
{\bf Keywords}: classical geometry, quantum geometry, classical reference frame, quantum reference frame, classical symmetry, quantum symmetry, Erlangen Program, Quantum Erlangen Program

\tableofcontents

\section{Introduction: What is the Erlangen Program?} \label{Sec.1}

The Erlangen Program was a research program in classical geometry, in the late nineteenth century, which was inaugurated by Felix Klein in 1872 \cite{ErlangenProgram_Klein_1872,ErlangenProgram_EMS_2015,Kisil:2010,Goenner:2015,ErlangenProgram_Encycolpedia_Srpinger}.
Its goal was to classify certain spaces, entirely in terms of their symmetries, in order to better understand the relations between them. 
Originally, the spaces considered were highly symmetric, possessing global symmetries that allowed them to be specified completely. 
Roughly speaking, the results of the program imply that
\begin{eqnarray} \label{Eq.1}
same \quad symmetries \iff same \quad geometry \, .
\end{eqnarray}
For example, suppose we have two geometries, but are not sure if they are equivalent. 
If we can prove that both spaces are invariant under the same group of global transformations, then they are the same abstract space. 

This is the main result of the original Erlangen Program, which proved that the `top down' approach to constructing metric spaces is completely equivalent to the `bottom up' approach used in conventional differential geometry. 
Recall that in the latter one starts with an abstract set and then performs a partial ordering, in order to build the elements of this set into a topological space. 
One then introduces congruences of differentiable curves, in order to define some notion of smoothness, generating a manifold. 
Finally, one endows the manifold with a metric by defining an inner product between the vectors tangent to these curves \cite{Frankel:1997ec,Nakahara:2003nw}. 
Then, and only then, can one consider the type of symmetry transformations that may leave this space invariant \cite{Jones:1998}. 

The Erlangen Program turns this approach on its head by starting with the symmetries and working backwards towards more primitive notions. 
Practically, it is extremely powerful and the results of the original program can be also generalised by considering local symmetries. 
In this case, the statement (\ref{Eq.1}) no longer holds, but its local version remains valid. 

For example, any two psuedo-Riemannian geometries with the same metric signature $(p,q)$ are locally isomorphic, obeying local $so(p,q)$ symmetry. 
But they may differ globally, possessing different topologies and curvature profiles. 
We may say that
\begin{eqnarray} \label{Eq.2}
same \quad local \quad symmetries \iff same \quad local \quad geometry \, .
\end{eqnarray}
This is still a very powerful result, with important implications for contemporary research, as it is sufficient to rule out any overly-simplistic claims about the role of local Poincar{\' e} invariance in quantum gravity. 
Strictly, there can be no such invariance, since imposing it is equivalent to imposing a classical Lorentzian background!

With that said, we will consider an important qualification of this statement in Sec. \ref{Sec.4}. 
We will show that it is possible for classical symmetries to be ``mathematically preserved but operationally broken'', in a quantum superposition of spacetimes. 
The essential point is that, despite superficial appearances to the contrary, this is not {\it really} the same as simply preserving classical symmetries, in the usual sense.

\section{What is the ``Quantum Erlangen Program?''} \label{Sec.2}

The intuitive idea behind this nascent research program comes from the following observations:
\begin{enumerate}

\item Classical geometries are completely specified, at least locally, by the group of local symmetry transformations.

\item The generators of classical local symmetries perform transformations between {\bf classical reference frames} (CRFs).

\item Though it is surprisingly difficult to define the term ``reference frame'' in abstract terms, we all know what it means intuitively.
One way to think of a CRF is as a weightless local coordinate system, whose origin follows an arbitrary worldline, and which is also free to rotate arbitrarily. 
The coordinates are not physical and cannot interact with any physical systems, including the background space in which they are defined 
(i.e., there is no gravitational back-reaction of the frame on the geometry).

\item No realistic observer is a weightless, non-interacting set of arbitrary coordinates! 
Real observers are embodied as physical systems, which, in principle, should be described by the laws of quantum mechanics. 
This is the fundamental insight of the {\bf quantum reference frame} (QRF) research program \cite{Giacomini:2017zju,Vanrietvelde:2018pgb,Krumm:2020fws,delaHamette:2021oex,Muller_Group_Website,Loveridge:2016tnh,Loveridge:2017pcv,Carette:2023wpz}.

\end{enumerate}

These observations prompt the following question: {\it If transformations between CRFs uniquely specify the local classical geometry, can transformations between QRFs be used to define locally {\bf quantised geometry}?}
\\ \\ \indent
The answer to this question depends, critically, on what exactly we mean by the term QRF, and how the transformations between different QRFs are constructed. 

\section{What is a ``Quantum Reference Frame?''} \label{Sec.3}

\subsection{The intuitive idea} \label{Sec.3.1}

Practically, we may identify a QRF with a {\it subsystem} of a closed physical system. 
For example, suppose that the system in question consists of three subsystems, each of which may perform measurements on the remaining two, but does not have access it its own `internal' degrees of freedom. 
Following the literature, we call these subsystems Alice, Bob and Charlie, QRF-$A$, QRF-$B$ and QRF-$C$, or simply $A$, $B$ and $C$, depending on how precise we wish to be. 
We will use the terms QRF-A and CRF-$A$, when we wish to distinguish clearly between a quantum and a classical reference frame, and simply $A$, or Alice, whenever our intended meaning is unambiguous.
As a first approximation, we treat each subsystem as a `particle', which is equivalent to considering only the degrees of freedom associated with its centre-of-mass, together with any net quantum mechanical spin it may possess. 
To make things even simpler, we will also neglect spin.

In this scenario, Alice can measure her relative distance to Bob, or to Charlie, but has no way to obtain information, in single-shot experiments, about the quantum mechanical spread of her own position. 
She can also measure the momenta of Bob and Charlie, relative to herself, but cannot obtain information about the shape of her own wave function in momentum space, without compiling statistics 

\subsection{The mainstream view} \label{Sec.3.2}

In fact, in the majority of the QRF literature, transformations between QRFs are assumed to be unitary, which means that the total information available to any subsystem is `perspective neutral' \cite{Giacomini:2017zju,Vanrietvelde:2018pgb,Krumm:2020fws,delaHamette:2021oex,Muller_Group_Website}.
\footnote{For an alternative to this view, much closer to that espoused here, see \cite{Loveridge:2016tnh,Loveridge:2017pcv,Carette:2023wpz} and related works.} 
This follows naturally from a {\it reduced quantisation} procedure, in which classical `ignorable coordinates' \cite{Kibble_Berkshire:2004} are ignored and only relative degrees of freedom are quantised. 
While this makes a good deal of intuitive sense, it assumes that classically ignorable degrees of freedom (e.g., the position and momentum of the {\it total} centre-of-mass of a closed system) remain ignorable in the quantum regime. 
Under this scheme, a three-particle system is described by only two relative displacements, or, equivalently, two relative momenta, while a four-particle system is described by only three relative displacements, and so on. 
Hence, under reduced quantisation, the designated observer does not possess a `wave function of its own', either in position or momentum space.
\footnote{Of course, even in canonical QM, a particle does not really possess a `wave function of its own' when it is entangled with another particle \cite{Rae:2002,Isham:1995} . But this is not what we mean here. In the bulk of the QRF literature, the subsystem designated as `the observer' does not possess any independent degrees of freedom and does not contribute to the total number of degrees of freedom of the system \cite{Giacomini:2017zju,Vanrietvelde:2018pgb,Krumm:2020fws,delaHamette:2021oex,Muller_Group_Website}. 
(Much like an ideal CRF.)}

\subsection{Our perspective} \label{Sec.3.3}

An alternative approach is to impose canonical {\it Dirac quantisation}, which is equivalent to adopting the perspective of an external CRF. 
This may seem misguided at first. 
After all, didn't we just argue, above, that all physical observers must obey the laws of quantum mechanics? 
We did. 
But in canonical QM, physical observers correspond to material quantum systems, situated in a {\it classical} background geometry. 

If such a classical background space exists -- and in canonical QM it surely does, at least formally \cite{Rae:2002,Isham:1995} -- then it is perfectly sharp, and not subject to quantum mechanical uncertainty. 
It can therefore be used to define CRFs, which, although they do not correspond to physical observers, remain valid as `objective' mathematical descriptions of physical systems. 

The key issue, therefore, is how to define a CRF-to-QRF transition? 
In other words, how do we go from the objective, complete, `God's eye view' of a CRF, to a the subjective, partial, and limited view of a physically embedded observer? 
The simple answer is that we take the a partial trace over the total Hilbert space of the system, tracing out the degrees of freedom corresponding the subsystem that has been designated as `the observer'. 
Considering the viewpoints of different subsystems then corresponds to taking different partial traces. 
Since each partial trace removes different information, there is no `perspective neutral' view \cite{Giacomini:2017zju,Vanrietvelde:2018pgb,Krumm:2020fws,delaHamette:2021oex,Muller_Group_Website}, and neither CRF-to-QRF, nor QRF-to-QRF transitions can be {\it unitary}.  

As a down-to-earth example of the philosophy underlying our approach, consider the colour of your eyes.
Are they brown, blue, or green? 
How do you know? 
Because you cannot see your own eyes, directly, you can only know their colour by interacting with another `particle': in this case, a mirror-particle or another human-particle who tells you what colour they are. 
There is no such thing as a perspective neutral view in which everyone knows their eye colour without looking in a mirror, or being told, or in which two pairs of eyes are shared between three people so that the question ``what colour eyes do {\it I} have?'' becomes meaningless, because eye colour is a purely relative concept.

\subsection{Relevance of the alternative view for quantum gravity} \label{Sec.3.4}

The example above highlights, in a non-technical way, the kind of absurd consequences that can result from following a reduced quantisation procedure. 
But the physical arguments in favour of this procedure are far from absurd, and it is by no means obvious that they are incorrect. 
In the classical regime, relative quantities are the only observable quantities: there is no way to detect the absolute inertial motion, or position, of the centre-of-mass of a closed system, and these concepts are not a legitimate part of physics! 
Why should it be any different in the quantum regime?

There are various ways to look at this question, which is of paramount importance to various issues in quantum gravity research. 
Ultimately, however, each concerns the question of what we mean when we say that something is `observable', so we will first analyse this concept in some depth.

In canonical QM, the term observable has a fixed and well-defined meaning. 
An observable is represented, mathematically, by a hermitian operator. 
The possible measured values of the observable are the (real) eigenvalues of this operator, which means that, in the canonical theory, observables can be measured directly, in single-shot observations \cite{Rae:2002,Isham:1995}. 
The relative displacement and relative momentum between two particles are good examples of observables. 
By contrast, there is no way in which the position or momentum of a single quantum particle can be measured, so these are {\it not} observables. 

We have already shown that, in classical physics, an unobservable quantity is unphysical, so it is tempting to think that the same must be true in quantum theory. 
But can we be sure that this is the case, without further examination of the problem? 
Just because there is a one-to-one correspondence between physical quantities and observables in classical physics doesn't mean, necessarily, that this one-to-one correspondence holds in the quantum regime. 
Put another way: all quantum observables are physical quantities, but are all physical quantities in quantum theory canonical observables? 

To help answer this question, it is instructive to first consider {\it classical} extended objects. 
Clearly, for a composite body, the centre-of-mass approximation is just that -- an approximation. 
The body possesses various `internal' degrees of freedom, i.e., the relative displacements and momenta between its constituent particle-like subsystems. 
It is therefore clear how to describe both point-particles and extended objects, dynamically, in the classical regime. 
A truly particle-like classical object has no internal degrees of freedom, and, since the absolute motion of its centre-of-mass is unobservable, a {\it single} classical particle possesses no degrees of freedom at all, as we have already noted. 

But extended classical objects {\it do} possess physical degrees of freedom \cite{Frankel:1997ec}. 
We are therefore prompted to ask another question, namely, whether a quantum `particle' is truly particle-like, in the same sense as a classical particle, or whether, because it may be associated with a wave function, it should instead be treated more like an extended object. 
If it is an extended object, either in position or momentum space -- and the uncertainty principle tells us that it is both \cite{Rae:2002,Isham:1995} -- then it is by no means clear that only two relative displacements, or two relative momenta, suffice to describe a three-particle system. 
It is at least conceivable that the spread of Alice's wave function, in position space, could contribute to the overall uncertainty in her observed values of Bob and Charlie's displacements (relative to herself). 
Likewise, the spread of her wave function in momentum space may introduce additional uncertainty into her measurements of their momenta. 

Beginning with this assumption, which is equivalent to assuming that a sharp spacetime background exists, imposing canonical Dirac quantisation from a CRF defined therein, and taking the partial trace over the subspace of the total Hilbert space associated with the designated observer (in this case Alice), a straightforward calculation yields \cite{Lake:2023nua,Lake:2020rwc,Lake:2023lvh} 
\begin{eqnarray} \label{Eq.1}
(\Delta_{\Psi}X_{B|A})^2 = (\Delta_{\psi}x_{B})^2 + (\Delta_{\phi}x_{A})^2  \, , \quad (\Delta_{\Psi}P_{B|A})^2 = (\Delta_{\psi}p_{B})^2 + (\Delta_{\phi}p_{A})^2  \, .
\end{eqnarray}
Here, we have considered a closed bipartite system, for simplicity: $\Psi_{AB}(x_A,x_B) = \phi_{A}(x_A)\psi_B(x_B-x_A)$ is the total wave function of Alice and Bob's joint state, whereas $\phi_{A}(x_A)$ and $\psi_{B}(x_B)$ are the individual wave functions of these subsystems in position space.
\footnote{Roughly speaking, $\psi_{B}(x_B)$ is the wave function that Bob ``would have'', if the observer Alice were not present.}
$X_{B|A} = x_B - x_A$ denotes the single {\it relative} displacement, which is the only position space observable.
Similarly, $\tilde{\Psi}_{AB}(p_A,p_B) = \tilde{\phi}_{A}(p_A)\tilde{\psi}_B(p_B-p_A)$ is the wave function of their joint state in momentum space and $P_{B|A} = p_B - p_A$ denotes the single, observable, relative momentum.

From this simple calculation, it is clear that, while $x_A$ and $x_B$ ($p_A$ and $p_B$) are {\it unobservable}, in the canonical sense, they nonetheless have {\it detectable} physical effects.
\footnote{There is some work for philosophers of science to do here! It would be helpful to have ready-made definitions of {\it observable} and {\it detectable} quantities, together with a list of clear delineating criteria. } 
Equation (\ref{Eq.1}) can be combined with the canonical uncertainty relations, $\Delta_{\phi}x_A\Delta_{\phi}p_A \geq \hbar/2$ and $\Delta_{\psi}x_B\Delta_{\psi}p_B \geq \hbar/2$, to give \cite{Lake:2023nua,Lake:2020rwc,Lake:2023lvh}  
\begin{eqnarray} \label{Eq.2}
\Delta_{\Psi}X_{B|A} \Delta_{\Psi}P_{B|A} \gtrsim \frac{\hbar}{2}\left[1 + \frac{(\Delta_{\phi}x_{A})^2}{\hbar^2}(\Delta_{\Psi}P_{B|A} )^2 + \frac{(\Delta_{\phi}p_{A})^2}{\hbar^2}(\Delta_{\Psi}X_{B|A} )^2 \right] \, . 
\end{eqnarray}
Using this relation, the physical existence of the $x_A$ and $x_B$ ($p_A$ and $p_B$) degrees of freedom can be {\it inferred}, using only statistical analyses of repeated $X_{B|A}$ and $P_{B|A}$ measurements.

We may now consider the problem we have working towards all along, namely, how is this relevant for quantum gravity? 
Naively, we expect spacetime to become `fuzzy' at the Planck scale. 
This Planck-scale fuzziness can be modelled, even in the non-relativistic regime, by delocalising the origin of our chosen CRF. 
Practically, this amounts to associating a canonical position eigenstate, $\ket{x}$, with each point in the classical background space, labelled by the coordinate $x$. 
We then `smear' this state by convolving it with a Planck-scale Gaussian wave function, yielding \cite{Lake:2018zeg,Lake:2019nmn,Lake:2021beh,Lake:2020chb}
\begin{eqnarray} \label{Eq.3}
x \mapsto \ket{x} \mapsto \ket{g_{x}} = \int g(x'-x)\ket{x'}{\rm d}^3x \, , \quad \braket{g_x|g_x} = \int |g(x'-x)|^2{\rm d}^3x = 1 \, .
\end{eqnarray}
By the same logic that leads to Eqs. (\ref{Eq.1}) and (\ref{Eq.2}), it is straightforward to show that the uncertainty relation obeyed by the `bipartite' state of Alice and Bob is now
\begin{eqnarray} \label{Eq.4}
\Delta_{\Psi}X_{B|A} \Delta_{\Psi}P_{B|A} \gtrsim \frac{\hbar}{2}\left[1 + \frac{(\Delta_{\phi}x'_{A})^2 + (\Delta_{g}x'_{A})^2}{\hbar^2}(\Delta_{\Psi}P_{B|A} )^2 + \frac{(\Delta_{\phi}p'_{A})^2 + (\Delta_{g}p'_{A})^2}{\hbar^2}(\Delta_{\Psi}X_{B|A} )^2 \right] \, . 
\end{eqnarray}
where
\begin{eqnarray} \label{Eq.5}
(\Delta_{\Psi}X_{B|A})^2 = (\Delta_{\psi}x'_{B})^2 + (\Delta_{\phi}x'_{A})^2 + (\Delta_{g}x'_{A})^2  \, , \quad (\Delta_{\Psi}P_{B|A})^2 = (\Delta_{\psi}p'_{B})^2 + (\Delta_{\phi}p'_{A})^2 + (\Delta_{g}p'_{A})^2  \, .
\end{eqnarray}

Actually, Alice and Bob's state is no longer bipartite, because neither can avoid interaction with the fluctuating spacetime they inhabit! 
It is not possible for either of them to measure their distance to an abstract spacetime point, and, to be absolutely clear, our model does {\it not} ascribe operational significance to spacetime points, which would violate, among other things, then tenets of Einstein's hole argument \cite{The_Hole_Argument_SEP}. 
But this does not stop fluctuations of the background geometry from affecting the statistical spreads of measurements they {\it can} make, on each other. 
As in our non-quantum gravity example, above, some degrees of freedom in the physical system do not correspond to observables. 
But they are nonetheless physical, as their presence has physical effects. 
These are detectable through the existence of modified statistics, which cannot be accounted for by canonical quantum theory.

This is, in fact, one of the main advantages of our model. 
By defining an alternative form of `quantum reference frame', in opposition to the mainstream view, we have constructed a simple model of Planck-scale spacetime that makes contact with existing quantum gravity phenomenology \cite{Maggiore:1993rv,Adler:1999bu,Scardigli:1999jh,Bolen:2004sq,Park:2007az,Bambi:2007ty,Hossenfelder:2012jw,Tawfik:2015rva}. 
When Alice's mass is large the spread of her centre-of-mass wave function can be made arbitrarily small, in both position space and velocity space, since $\Delta_{\phi} v_A \gtrsim \hbar/(2m_A\Delta_{\phi} x_A)$. 
In this limit, her QRF becomes {\it effectively} classical, and at least one term in the leading order corrections to canonical uncertainty relation is provided by the Planck-scale fluctuations of spacetime! 
Setting $\Delta_{g}x'_{A}$ to be of the order of the Planck length and $\Delta_{g}p'_{A}$ to be of the order of the de Sitter momentum,
\begin{eqnarray} \label{Eq.6}
\Delta_{g}x'_{A} \simeq \sqrt{\hbar G/c^3} \simeq 10^{-33} \, {\rm cm} \, , \quad \Delta_{g}p'_{A} \simeq \hbar\sqrt{\Lambda/3} \simeq 10^{-56} \, {\rm g \, cm \, s^{-1}} \, , 
\end{eqnarray}
then recovers the extended generalised uncertainty principle (EGUP) \cite{Bolen:2004sq,Park:2007az,Bambi:2007ty}, a modified uncertainty relation that has been predicted, on model-independent grounds, in the quantum gravity literature: 
\begin{eqnarray} \label{Eq.7}
\Delta_{\Psi}X_{B|A} \Delta_{\Psi}P_{B|A} \gtrsim \frac{\hbar}{2}\left[1 + \frac{2G}{\hbar c^3}(\Delta_{\Psi}P_{B|A} )^2 + \eta_0\Lambda(\Delta_{\Psi}X_{B|A} )^2 \right] \, . 
\end{eqnarray}
In this expression $\Lambda \simeq 10^{-56}$ ${\rm cm}^{-1}$ is the cosmological constant \cite{Martin:2012bt} and $\eta_0$ is a numerical constant. 

We can even go slightly further with this. 
The model outlined can actually be summarised in a single line, since it is equivalent to imposing the modified de Broglie relation   
\begin{eqnarray} \label{Eq.8}
p' = \hbar k + \beta(k'-k) \, ,
\end{eqnarray}
on all {\it material} quantum particles, where $\beta \simeq \Delta_{g}x'_{A} \Delta_{g}p'_{A} \simeq \hbar \times 10^{-61}$ \cite{Lake:2018zeg,Lake:2019nmn,Lake:2021beh,Lake:2020chb}. 
Here, $p = \hbar k$ is the canonical quantised momentum and the non-canonical term $\beta(k'-k)$ can be thought of, heuristically, as an additional momentum `kick', generated by quantum fluctuations of the background geometry. 

The key point is that $p'$ is the only {\it measurable} momentum, whereas the unprimed degrees of freedom are integrated out in the derivation of the EGUP. 
This is equivalent to performing a partial trace over the subsystem associated with quantum state of the geometry, and this is necessary because, as we have already stressed, no physical measurements can be performed on abstract spacetime points \cite{The_Hole_Argument_SEP}. 
The degrees of freedom associated with the quantised background are therefore not observables, in the usual sense of this term, but they are still physical degrees of freedom. 
Their existence is {\it detectable}, via statistical inference, and the modified statistics they create are manifested in the existence of generalised uncertainty relations, which are the hallmark, or tell-tale signature, of the quantum gravity regime \cite{Maggiore:1993rv,Adler:1999bu,Scardigli:1999jh,Bolen:2004sq,Park:2007az,Bambi:2007ty,Hossenfelder:2012jw,Tawfik:2015rva,Lake:2018zeg,Lake:2019nmn,Lake:2021beh,Lake:2020chb}.

Taking the limit $\beta \rightarrow 0$ of our quantum geometry model recovers canonical QM, whereas taking the $\beta \rightarrow \hbar$ limit recovers the alternative definition of a material QRF, discussed in Sec. \ref{Sec.3.3}. 
However, we see no way to achieve these, or similar results, by invoking unitary transformations between QRFs, as in the mainstream approaches to this subject \cite{Giacomini:2017zju,Vanrietvelde:2018pgb,Krumm:2020fws,delaHamette:2021oex,Muller_Group_Website}. 

\section{Towards the QEP: spacetime symmetries -- preserved, broken, and ``smeared''} \label{Sec.4} 

In the previous section, we obtained a naive model of Planck-scale quantised geometry by `smearing' the coordinate origins of sharp classical frames of reference. 
We can now combine this result with the logic of the original Erlangen program. 
As a concrete example, consider the non-relativistic limit represented by the EGUP (\ref{Eq.7}), which represents first-order quantum gravity corrections to ordinary Euclidean space. 

The symmetries of Euclidean space are well known, and it well known that they give rise, rather directly, to the uncertainty relations of canonical QM. 
The link between symmetries and uncertainties is provided by the Schr{\" o}dinger-Robertson relation, which holds for any hermitian operators $\widehat{O}_1$ and $\widehat{O}_2$ \cite{Rae:2002,Isham:1995}: 
\begin{eqnarray} \label{Eq.9}
\Delta_{\psi}O_1 \Delta_{\psi}O_2 \geq \frac{\hbar}{2} |\braket{\psi | [\widehat{O}_1,\widehat{O}_2] | \psi}| \, .
\end{eqnarray}
For example, the canonical position-momentum commutator is obtained by identifying the wave number operator, $\widehat{k} = \widehat{p}/\hbar$, with the generator of spatial translations, giving $[\widehat{x},\widehat{k}] = i$. 
The Heisenberg uncertainty relation $\Delta_{\psi}x\Delta_{\psi}p \geq \hbar/2$ therefore follows, directly, from the homogeneity (i.e., translation invariance) of Euclidean geometry. 
Likewise, the canonical angular momentum operators can be identified with the generators of rotations (up to a factor of $\hbar$) and their algebra is a direct consequence of rotational invariance. 

How, then, are these symmetries changed at the Planck scale, according to our model? 
Simply substituting the generalised position and momentum operators, $\widehat{X}_{B|A}$ and $\widehat{P}_{B|A}$, into (\ref{Eq.9}), gives a surprising answer: it {\it seems} to imply that they are not changed at all! 
The generalised commutator is simply 
\begin{eqnarray} \label{Eq.10}
[\widehat{X}_{B|A},\widehat{P}_{B|A}] = i(\hbar+\beta) \, , 
\end{eqnarray}
giving
\begin{eqnarray} \label{Eq.11}
\Delta_{\psi}X_{B|A}\Delta_{\psi}P_{B|A} \geq \frac{(\hbar+\beta)}{2} \, .
\end{eqnarray}
The revised Schr{\" o}dinger-Robertson bound, $(\hbar+\beta)/2$, which differs from the canonical bound only by one part in $10^{61}$, represents the lowest possible value of the EGUP bound (\ref{Eq.7}). 
It is therefore compatible with generalised uncertainty relations, induced by Planck-scale fluctuations of the background geometry \cite{Lake:2018zeg,Lake:2019nmn,Lake:2021beh,Lake:2020chb}.
But the commutator (\ref{Eq.10}) is still of canonical form, which implies that it is {\it still} a result of an {\it exact} translational symmetry! 

How can we get {\it different} uncertainty relations from the {\it same} symmetry? 
(Don't the results of the Erlangen Program expressly forbid it?)
The answer to this question is subtle but cuts to the heart of the distinction between {\it observables} and what have called {\it detectables} (for want of a better term) in Sec. \ref{Sec.3}. 
At first glance, it looks like we have just rescaled $\hbar$ and left the underlying structure of canonical QM unchanged. 
But this is not so. 
We may `unpack' the algebra (\ref{Eq.10}) into its constituent subalgebras, namely $[\widehat{x}_{A/B},\widehat{p}_{A/B}] = i\hbar$ and $[(\widehat{x'-x})_{A/B},(\widehat{p'-p})_{A/B}] = i\beta$, where the latter follows from a rearrangement of the modified de Broglie relation (\ref{Eq.8}): $(\widehat{p'-p}) = \beta(\widehat{k'-k})$, with $\widehat{p}=\hbar \widehat{k}$ \cite{Lake:2018zeg,Lake:2019nmn,Lake:2021beh,Lake:2020chb}.
\footnote{Note that, here, ``$A/B$'' means ``$A$ and/or $B$''. This should not be confused with the notation $B|A$, used in (\ref{Eq.7}), which means ``the state of $B$, from the perspective of $A$''. } 

The physical situation then becomes much clearer. 
What we have done, in fact, is to enlarge the Hilbert space of Alice and Bob's original bipartite state, to include degrees of freedom associated with fluctuations of the background. 
These fluctuations can be viewed as small perturbations away from the canonical state, so that we must now include the both the $p_{B|A}$ and $(p'-p)_{B|A}$ degrees of freedom -- or, equivalently, the $x_{B|A}$ and $(x'-x)_{B|A}$ degrees of freedom -- in our physical description of the system. 
For the momenta, we have $p_{B|A} = p_{B}-p_{A}$ and $(p'-p)_{B|A} = (p'-p)_{B} - (p'-p)_{A}$. 
Since the modified de Broglie relation (\ref{Eq.8}) is linear in both $\widehat{k}$ and $(\widehat{k}'-\widehat{k})$, it is clear that we have imposed translational symmetry on {\it both} canonical Euclidean space, labelled by the coordinates $x_{A/B}$, {\it and} `fluctuation space', labelled by coordinates $(x'-x)_{A/B}$. 

However, these symmetries do not generate equal contributions to the physical momentum of the system. 
The translation invariance of the classical background space generates the canonical contribution, $\widehat{p}=\hbar \widehat{k}$, whereas the translation invariance of fluctuation space contributes an additional `perturbation', $(\widehat{p'-p}) = \beta (\widehat{k'-k})$. 
Furthermore, since the only {\it measurable} momentum is the resultant momentum $\widehat{p}' = \hbar\widehat{k} + \beta (\widehat{k'-k})$ -- including both canonical and perturbative contributions -- the mathematical structure of our theory also imposes translational invariance on the {\it space of measurable values}. 
That is, the abstract space labelled by the coordinates $\widehat{x}'_{B|A} = \widehat{x}'_{B}-\widehat{x}'_{A}$. 
It is this final symmetry that accounts, ultimately, for the $(\hbar + \beta)$ scaling of the modified commutator (\ref{Eq.10}) \cite{Lake:2018zeg,Lake:2019nmn,Lake:2021beh,Lake:2020chb}.   

The analysis above highlights the differing roles of {\it detectables} and {\it observables}. 
In this model, classical Euclidean space does not physically exist, but the Hilbert space of the composite matter-plus-geometry system still contains subspaces labelled by the position variables $x_{A/B}$. 
In canonical QM, these variables {\it would} represent the position of a material quantum particle in a fixed, classical, Euclidean geometry.
In our model, a partial trace is performed over these subspaces, reflecting the fact that no `unperturbed' position value can ever be physically measured. 
In summary, both $x_{B|A}$ and $(x'-x)_{B|A}$ label physical degrees of freedom, but are {\it not} observable quantities. 
On the other hand, $x'_{B|A}$ {\it is} an observable, but the primed coordinates do not label actual positions in a {\it fixed} physical geometry. 
There is no fixed physical geometry and the would-be experimenter has no control over fluctuations in the space of perturbations, labelled by $(x'-x)_{B|A}$. 
All he or she can observe is the net result of {\it both} contributions -- canonical and non-canonical --  to the overall uncertainty in the position or momentum of a material quantum particle.  

To see what this means, practically, for the resolution of spacetime symmetries, consider the following thought experiment. 
Let us imagine an experimenter who prepares an array of heavy particles, with fixed inter-particle distances, at an initial time $t=0$. 
We assume that the particles all have the same mass and that this is large enough to render the canonical position uncertainty sub-Planckian, $\Delta x \simeq \hbar/(M\Delta p) \ll 10^{-33}$ ${\rm cm}$. 
This can be achieved, for reasonable values of the momentum uncertainty (i.e., those giving rise to an undetectable uncertainty in velocity space, $\Delta v = \Delta p/M \simeq 0$), for masses far in excess of the Planck mass. 
Under these conditions, the masses are effectively classical and the leading order contributions to the position uncertainty come from the Planck-scale fluctuations of the spacetime geometry. 
\footnote{This result is somewhat surprising, when first encountered, as it implies that macroscopic objects are the most effective probes of the quantum gravity regime! But a moment's reflection reveals that this statement is true - or, more specifically, that it would be true -- {\it if} our instruments were sensitive enough to measure Planck-scale effects. If this were the case, we could prepare a massive object which, according to canonical QM, should have a sub-Planckian uncertainty associated with the position of its centre-of-mass: any Planck-scale uncertainty then observed would evidence of the quantum gravity regime. Practical limitations, not theory, prevent this strategy from working, but we will consider a hypothetical situation in which our experimenter is able to `zoom in all the way', from the atomic scale to the Planck scale.} 

Our experimenter then tries to investigate the symmetry properties of spacetime, at extremely small scales, by rigidly translating the lattice of heavy masses by some small distance $a$. 
In canonical QM this operation would be represented, mathematically, by an operator of the form $\widehat{U}(a) = \exp(i\widehat{p}a)$, where $\widehat{p} = \hbar\widehat{k}$ is a canonical momentum operator. 
In our model, however, the canonical momentum operator does not exist, and must be replaced by its `perturbed' counterpart $\widehat{P} = \hbar\widehat{k} + \beta(\widehat{k'-k})$. 
(As in Sec. \ref{Sec.3}, we have used the upper case $\widehat{P}$ to denote generalised momentum operators, whose eigenvalues are given by the net momentum $p'$ (\ref{Eq.8}).)
The resulting unitary $\widehat{\mathcal{U}}(a) = \exp(i\widehat{P}a)$, where $\widehat{P} = \hbar\widehat{k} + \beta\widehat{(\widehat{k}'-\widehat{k})}$, represents the earnest attempt of our experimenter to rigidly translate the lattice by the distance $a$, but he or she {\it cannot} be successful in this task. 
The rather bad drawing below illustrates, schematically, what happens instead. 
(See Fig. 1.)   
\begin{figure}[h] \label{Fig.1}
\begin{center}
\includegraphics[width=16cm]{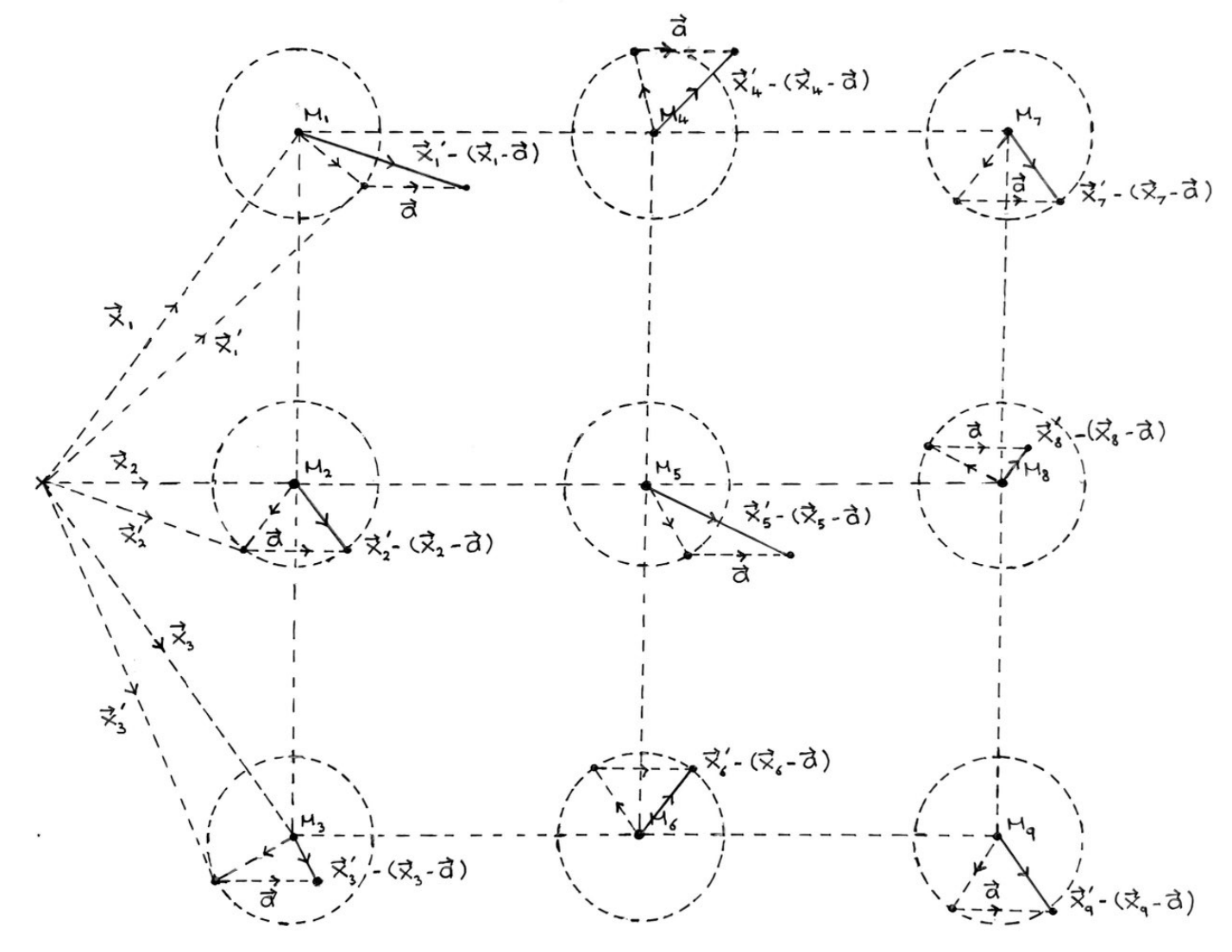}
\caption{Schematic representation of an experiment designed to probe the translational invariance of space at small scale. (See text for details.) Taken from \cite{Lake:2023nua}, with permission.}
\end{center}
\end{figure}

In Fig. 1, the initial positions of nine masses, labelled $M_{i}$ with $i = 1,2, \dots 9$, are represented by black dots and the canonical quantum uncertainty in their positions is assumed to be negligible. 
Their initial positions, relative to the experimenter, are labelled by the coordinates $x_i$, and the initial displacement vectors are depicted as long dashed arrows, ending on the dots. 
The initial positional of the experimenter is represented by a cross, and, for simplicity, we have neglected all contributions to the uncertainty in this value. 
(In a more careful analysis, these must certainly be accounted for. However, for an effectively classical observer, the net uncertainty in position will also be of the order of the Planck length.) 
The circles drawn with dashed lines surrounding each dot represent the width of the Planck-sized Gaussian function, $|g(x'-x)|^2$, introduced in Sec. \ref{Sec.3.4}. 
Their radii, of approximately one Planck length, give the characteristic scale of fluctuations induced by the geometry, in the vicinity of each mass. 
This is the characteristic length of the vector $(x'_i-x_i)$, which, however, is {\it randomly oriented}. 

What this diagram represents is the fact the smeared-space translation operator, $\widehat{\mathcal{U}}(a) = \exp(i\widehat{P}a)$, translates the {\it perturbed} position vector, $x'_i = x_i + (x'_i-x_i)$, by $a$. 
This should be contrasted with action of the canonical translation operator, which translates $x_i$ by $a$, since no background-induced perturbation exists in the canonical theory. 
When $|a| \gg |x'_i-x_i| \simeq 10^{-33}$ ${\rm cm}$ -- that is, when the distance through which the experimenter tries to translate the lattice is much larger than the Planck-scale perturbation -- the angular deflection that the perturbation induces is unmeasurably small. 
In this limit, the perturbations are effectively `washed out' and we recover the {\it appearance} of a rigid translation of the lattice. 
We therefore recover the appearance of smooth homogeneous space \cite{Lake:2023nua}. 

As we zoom in, taking $|a|$ closer and closer to the Planck length, the appearance of translation invariance, manifested at super-Planckian scales, breaks down. 
Our attempt to translate the lattice rigidly -- that is, without altering the relative displacements of the nine masses -- fails miserably, and its crystalline structure is randomly disordered. 
This is what we mean when we say that, in our quantum geometry model, classical symmetries are ``mathematically preserved but operationally broken'', as alluded to in the Introduction.

We stress, once more, that translational symmetry is not broken, in the usual sense of this term. 
For example, as in when we talk about the breaking of $SU(2) \times U(1)$ symmetry in electro-weak theory, where the symmetry is reduced to the $SU(2)$ and $U(1)$ subgroups associated with spin and electromagnetism \cite{Strocchi_Book}. 
But nor is it `unbroken', or `preserved', in the usual sense. 
At present, perhaps we lack the appropriate terminology to describe exactly what we mean. 
As a stop-gap, and, having introduced the conceptual framework at a somewhat technical level, we propose the term `smeared symmetry' for the phenomenon we describe, using the formalism outlined above. 

The analysis for rotational symmetry is more complex, but follows the same principles, and it is possible to show that this too can be smeared, generating modified uncertainty relations for angular momentum \cite{Lake:2019nmn}. 
Similar analyses can also be performed for spacetime symmetries, in the relativistic limit, leading to smeared Lorentz transformations, and, interestingly, as notion of smeared quantum mechanical spin \cite{Lake:2019nmn,Lake:2021beh}.

One final point we wish to make is that the unitary operator $\widehat{\mathcal{U}}(a) = \exp(i\widehat{P}a)$ should not be interpreted as representing a transformation between QRFs. 
Practically, the experimenter {\it cannot} `adopt the perspective' of the masses he or she measures, or even that of any one particular mass. 
The construction of the experimenter's frame involves a trace over the experimenter's internal degrees of freedom -- which are physical, but unmeasurable (to him or herself). 
These degrees of freedom are, however, accessible to other `observers', including the masses $M_i$. 
(Recall the example of eye colour, discussed in Sec. \ref{Sec.3.3}.)

To jump between QRFs requires the initial observer to input information that he or she does not have direct access to, in single-shot experiments, and to simultaneously remove information that he or she may in fact possess -- to `un-trace' one subspace of the total Hilbert space of the system, while tracing out another. 
What this means, in operational terms, is an open question. 
It is certainly possible, as we have shown, to obtain limited information from {\it detectable} but {\it unobservable} degrees of freedom. 
It may be the case that sufficient information can be gained, through the compilation of statistics, to effectively specify a `perspective neutral' view \cite{Giacomini:2017zju,Vanrietvelde:2018pgb,Krumm:2020fws,delaHamette:2021oex,Muller_Group_Website}, but this remains to be seen. 

Arguably, however, it may be more interesting if such a view is unattainable, even theoretically. 
In this case, the necessarily {\it approximate} nature of QFR-to-QFR transitions -- or, in everyday speech, the impossibility of ever truly attaining the perspective of another entity, no matter how hard one tries -- is intimately related, in the quantum gravity regime, to the unobservability of abstract spacetime points. 
No one believes that quantum fluctuations of the spacetime geometry should leave the material objects within it unaffected, but this does not mean that we can obtain complete information about these fluctuations, through observations of material objects alone. 
Yet, since these are the only objects on which we can perform physical measurements, this is the best we can do. 
Potentially, this has huge implications for the thermodynamics of spacetime, the (accessible) information content of the universe, and any spacetime phenomenon involving an ``information loss paradox'' \cite{Mathur:2009hf}. 

It may be hoped -- though, at this stage, it really is just a hope, and I frankly admit that I have no idea how to implement it in practice -- that a complete specification of an {\it optimal} QRF-to-QFT transition, one that preserves the maximal amount of information available to observer, may be used to classify the {\it quantised} geometry in which this transition occurs. 
This would be, in my view, the closest we can get to the results of the original Erlangen program, in the quantum gravity regime. 
This is the ultimate goal of our would-be research program: the {\it QEP}.  

\section{Conclusions} \label{Sec.5}

We have introduced an alternative definition of the term `quantum reference frame', which differs somewhat from the mainstream view, whereby QRF-to-QRF transitions are represented by unitary operators \cite{Giacomini:2017zju,Vanrietvelde:2018pgb,Krumm:2020fws,delaHamette:2021oex,Muller_Group_Website}. 
Our approach is motivated by the idea that at least {\it some} of the physical degrees of freedom, in any quantum mechanical system, may not correspond to {\it observables}. 
We call such degrees of freedom, which have physical effects but which cannot be measured directly in single-shot experiments, {\it detectables}. 
In this preliminary analysis, we have shown that the existence of detectable but unobservable degrees of freedom can always be {\it inferred}, by analysing the statistics obtained from repeated measurements of observable quantities alone. 

Although these conclusions are very general, our primary interest is their significance for quantum gravity research, and, in particular, for the symmetries of quantised spacetime. 
Indeed, an immediate consequence of such modified statistics is the existence of generalised uncertainty relations, of the kind predicted, generically, by phenomenological arguments in the quantum gravity literature \cite{Maggiore:1993rv,Adler:1999bu,Scardigli:1999jh,Bolen:2004sq,Park:2007az,Bambi:2007ty,Hossenfelder:2012jw,Tawfik:2015rva,Lake:2018zeg,Lake:2019nmn,Lake:2021beh,Lake:2020chb}. 
We introduced a simple model in which Planck-scale fluctuations of the background geometry disturb the relative displacements and momenta of material quantum particles, and used this to derive one of the most well-studied of these relations, the extended generalised uncertainty principle (EGUP) \cite{Bolen:2004sq,Park:2007az,Bambi:2007ty}. 

Central to the construction of our model is the fact that quantised fluctuations of the background have physical effects, despite the fact that no operational significance can be ascribed to spacetime points: that is, one cannot perform physical measurements on spacetime points {\it per se}, only on physical bodies separated by spacetime intervals \cite{The_Hole_Argument_SEP}. 
The degrees of freedom associated with these fluctuations are, therefore, paradigm examples of what we mean by {\it detectable but unobservable} degrees of freedom.    

Though our model may be somewhat naive, it has a straightforward physical interpretation. 
It may be viewed as a consequence of {\it delocalising} classical frames of reference, by `smearing' their coordinate origins over volumes comparable to the Planck volume \cite{Lake:2018zeg,Lake:2019nmn,Lake:2021beh,Lake:2020chb}. 
This view is, in effect, the application of our alternative definition of a QRF to the degrees of freedom responsible for fluctuations of the spacetime \cite{Lake:2023nua,Lake:2020rwc}. 

Finally, we attempted to combine this operational viewpoint with geometric and symmetry arguments. 
Put simply, the existence of generalised uncertainty relations requires the existence of modified commutation relations, and these relate, in turn, to the symmetry properties of the underlying space. 
We have shown that, in our model, the classical spacetime symmetries are ``mathematically preserved but operationally broken'', while QRF-to-QRF transitions {\it cannot} be unitary, since no two observers have access to exactly the same information \cite{Lake:2023nua}. 

Even if such a perspective-neutral view is possible, in principle, for material quantum systems in a fixed classical background \cite{Giacomini:2017zju,Vanrietvelde:2018pgb,Krumm:2020fws,delaHamette:2021oex,Muller_Group_Website}, the random nature of fluctuations in a quantised spacetime, over which material systems have no physical control, generates irremovable differences between measurements of a single relative variable, by two different observers. 
The best one can do is to average over the effects of these fluctuations, as manifested in repeated measurements of observable quantities, which is equivalent to taking the partial trace over the subsystem associated with the quantised background in the composite matter-plus-geometry system. 

It may be hoped that, despite such limitations, we may still be able to construct a non-unitary operator, corresponding to an optimal QRF-to-QRF transition, which preserves the maximum amount of information available to the designated observer. 
This would be the closest we could get, operationally, to the transformations between CRFs, which can be used to classify the symmetries of classical geometries according to the Erlangen Program of the nineteenth century \cite{ErlangenProgram_Klein_1872,ErlangenProgram_EMS_2015,Kisil:2010,Goenner:2015,ErlangenProgram_Encycolpedia_Srpinger}. 
It would, therefore, be reasonable to interpret such QRF-to-QRF transformations as an analogous classification of {\it quantised geometry} -- in effect, realising a twenty-first century update of Klein's original vision -- a {\it Quantum Erlangen Program}.


\end{document}